\documentstyle [epsfig,emulateapj] {article}

\newcommand{\be}{\begin{eqnarray}}
\newcommand{\ee}{\end{eqnarray}}
\newcommand{\bi}{\begin{itemize}}
\newcommand{\ei}{\end{itemize}}
\def\lsim{\mathrel{\rlap{\lower3pt\hbox{\hskip1pt$\sim$}}
    \raise1pt\hbox{$<$}}} %less than or approx. symbol
\def\gsim{\mathrel{\rlap{\lower3pt\hbox{\hskip1pt$\sim$}}
    \raise1pt\hbox{$>$}}} %greater than or approx. symbol
\newcommand{\msun}{\mbox{~$M_\odot$}}

\def\conv{{\rm conv}}
\def\dynamo{{\rm dynamo}}

\def\ms{{\rm ms}}
\def\s{{\rm s}}
\def\G{{\rm G}}
\def\H{{\rm H}}
\def\NS{{\rm NS}}

\begin{document}

%\rightline{KIAS-00-xxx}
%\rightline{SUNY-NTG-00-8}
%\leftline{Draft Version \today}

\title{On Companion-Induced Off-Center Supernova-Like Explosions}
\author{Chang-Hwan Lee$^{a,b}$, Insu Yi$^{b}$, and  Hyun Kyu Lee$^{c}$}
\affil{
  a) Dept. of Physics \& Astronomy, SUNY,
        Stony Brook, New York 11794, USA\\
  b) School of Physics, Korea Institute for Advanced Study, Seoul 130-012,
  Korea\\
  c) Dept. of Physics, Hanyang University, Seoul 133-791, Korea}

%---------------------------------------------------------------------

\begin{abstract}
We suggest that a neutron star with a strong magnetic field, spiraling
into the envelope of a companion star, can generate a ``companion
induced SN-like off-center explosion". The strongly magnetized neutron
star (``magnetar") is born in a supernova explosion before entering into
an expanding envelope of a supergiant companion. If the neutron star
collapses into a black hole via the hypercritical accretion during the
spiral-in phase, a rapidly rotating black hole with a strong magnetic field
at the horizon results. The Blandford-Znajek power is sufficient to power
a supernova-like event with the center of explosion displaced from the
companion core. The companion core, after explosion, evolves into a
C/O-white dwarf or a neutron star with a second explosion.
The detection of highly eccentric black-hole, C/O-white dwarf
binaries or the double explosion structures
in the supernova remnants could be an evidence of the proposed scenario.
\end{abstract}

%---------------------------------------------------------------------
\section{Introduction}
It has recently been reported that neutron stars with large luminosities
could be powered by ultra-strong magnetic fields $\sim 10^{14}G$ ("magnetars")
rather than rapid rotation as in rotation-powered pulsars
(e.g. Kouveliotou et al. 1998, 1999). These "magnetars" are
believed to be produced in supernova explosions or accretion induced collapses
(Duncan \& Thompson 1992).

On the other hand, the Blandford-Znajek process has been suggested as a
powerful source for a variety of very energetic events ranging from
active galactic nuclei (Blandford \& Znajek 1977) to gamma-ray bursts
(e.g. Lee et al. 2000ab and references therein).
The extremely high luminosities or power outputs generally require
ultra-strong magnetic fields around rapidly rotating black holes.
The origin of such strong magnetic fields around the rapidly spinning black
holes has not been clearly understood. It is interesting
to point out that if magnetars directly collapse to black holes via rapid
accretion, the resulting black holes could spin rapidly with magnetic fields
left at the horizon, which are large enough to make the Blandford-Znajek power
interesting for very energetic events.

Based on this possibility, we propose that the hypercritical accretion onto
a neutron star with a very strong magnetic field
can produce a rapidly spinning black hole with an ultra-strong field near
the horizon. The hypercritical accretion has been invoked for the common
envelope phase evolution of the massive stellar binaries (Bethe \& Brown
1998). Unless the strong
magnetic fields decay within time scales shorter than the duration of the
accretion phase, $\sim$ 1 year,
it is highly likely that the black holes with very strong fields could
be produced.
The Blandford-Znajek power is sufficient to power
a supernova-like event with the center of explosion displaced from the
companion core. The companion core, after explosion, evolves into a
C/O-white dwarf or a neutron star with a second explosion.
The detection of highly eccentric black-hole, C/O-white dwarf
binaries or the double explosion structure
in the supernova remnants, with two different centers of
explosion, could be an evidence of the proposed scenario.
Especially, in the later case, the neutron star can have double kicks,
resulting in very high kick velocities, $>1000$ km s$^{-1}$.
If the spiral-in
of the neutron star leads to the merger with the companion core, the
explosion at the core would appear as an explosion from a single star.
The neutron star-neutron star binary system is expected when the companion
collapses before significant accretion of the neutron star occurs.

%-----------------------------------------------------------------------------
\section{Magnetar Formation and Ultra-Strong Magnetic Fields}

The NSs right after core collapse are hot and convective throughout most of the
stellar interior due to a large neutrino flux. Duncan \& Thompson (1992)
and Thompson \& Duncan (1993)
describe the most likely outcome of such hot neutron stars. Right after the
collapse, there exists an intense neutrino-driven convective phase
with the overturn time scale of convective cell,
$\tau_{\conv}\sim 1$ ms at the base of the convection zone and
$\tau_{\conv}\sim 0.1$ ms at the neutrinosphere.
In this case, if the neutron star spin frequency is
    \be
    \Omega\gsim\Omega_{\dynamo}\sim 2\pi\tau_ {\conv}^{-1}\sim 6\times 10^3
    {\rm s^{-1}},
    \ee
the amplification of of a magnetic field by helical
motions is not suppressed by turbulent diffusion, and the $\alpha\Omega$-type
dynamo could work efficiently.
They found that the saturation field strength of
$B_{\rm sat} \sim 10^{16}-10^{17} \G$
could be reached within $\sim 10-100$ ms.
Therefore, post-collapse NSs with $P\lsim 1\ms$, within the first few seconds,
can build up very strong large scale magnetic fields $\ge 10^{15}\G$
through exponential growth. In this case, the resulting ultra-strong
field is independent of the fossil field strength. For strong seed fields,
the small-scale fluid motion, which is required for ohmic dissipation of
magnetic energy, is suppressed by the magnetic tension.
The convective motion ceases when the neutrino-driven convection becomes
unavailable.
Such a strong field pulsar with slow spin ("magnetar", Duncan \& Thompson
1992) has recently been discovered,
which strongly supports the basic idea of creation of a ultrastrong
field pulsar (Kouvelioutou et al. 1998, 1999).

For NSs spinning with $\Omega\lsim \Omega_{\dynamo}$ or $P\gsim 1\ms$,
the differential rotation (i.e. $d\Omega/dR\ne 0$) could amplify the
seed poloidal fields by wrapping them up into strong toroidal fields
(Kluzniak \& Ruderman 1998).
For an initial poloidal seed field with strength $B_p\sim 10^9\G$ which
would result from flux freezing of a field of $\lsim 3\times 10^4\G$
in the pre-collapse core,
is amplified to the strong toroidal field of $B_{\phi}\sim 10^{17}\G$ after
$\sim (1/2\pi)(B_{\phi}/B_p)\sim 10^7$ rotations or time
$\tau \sim B_{\phi}/B_p\Omega\sim 10^4 P\ \s$
with $P$ in $\ms$.
The amplified field becomes buoyant after reaching the critical field
strength of $B_b\sim 10^{17}\G$. Since the total energy included in the
buoyant fluid cells is $\sim \frac{1}{8\pi} B_b^2 (\frac{4\pi R^3}{3}) \eta
\sim  3\times 10^{51}\eta$ erg, where $\eta\sim 0.2$ is the
fractional volume of the buoyant elements, the repeated build-up and buoyant
loss will occur for a number of times determined as
   \be
   N_{\rm buoyant \ loss}
   \sim \frac{(1/2)I\Omega^2}{(1/8\pi)(B_b^2)(4\pi R^3/3)\eta}\sim
   \frac{2}{\eta}\sim 10,
   \ee
where $I\approx \frac 25 M R^2$ with neutron star mass $M=1.5\msun$
and neutron star radius $R=10$ km.
This process will continue until the existing
poloidal seed field is exhausted as the poloidal field is not
re-generated in this non-dynamo amplification process.
The buoyantly rising field could dissipate on a time scale
   \be
   \tau_{\rm dissipation} \sim (4\pi \rho)^{1/2}
   \frac{R}{B_b} \sim 10^{-3}\ \s,
   \ee
which controls the overall time scale.
Therefore, we conclude that essentially all NSs with $P\sim$ a few ms
could achieve strong fields through either dynamo or differential rotation.

Once strong fields are created, rapid spin-down of NSs is inevitable.
The electromagnetic dipole radiation luminosity is
\be
L_{\rm EM}\sim \frac{2}{3 c^3} B^2 R^6\Omega^4 =2\times 10^{50} B_{15}^2
\Omega_4^4 \ {\rm erg\ \s^{-1}}
\ee
which gives the characteristic electromagnetic spin-down time scale
\be
\tau_{\rm EM}\sim \frac{I\Omega^2}{2 L_{\rm EM}}\sim 3\times 10^2
\frac{1}{B_{15}^{2} \Omega_4^{2}} \ \s
\ee
where $B_{15}=B/10^{15}\G$ and $\Omega_4=\Omega/10^4 \s^{-1}$.
Therefore, all MSPs with ms periods are likely to spin-down to near the
pulsar deathline, $P\sim 70 B_{15}^{1/2}\s$, on a time scale
of $\sim 10^5 B_{15}^{-1}$ yr (e.g. Ritchings 1976).
That is, these initially fast spinning
MSPs would not be detected as MSPs unless they are recycled through the
low mass X-ray binary stage, which is not promising (Yi \& Grindlay 1998).
Therefore, before the common envelope phase begins, the neutron stars are
likely to be spinning with spin periods very similar to those of the
magnetars and anomalous X-ray pulsars (e.g. Mereghetti 1999, Kouvelioutou 1998, 1999).
That is, the initial conditions for the neutron stars which spiral-in during
the common envelope phase are likely to be a slow spin and a high magnetic
field.

%----------------------------------------------------------------------------
\section{Hypercritical Accretion}

Bethe \& Brown (1998) suggested that the hypercritical accretion can occur
in the massive binary systems, in which both stars can go into supernova
leaving the compact cores as remnants. In their scenario, the more massive star
\footnote{The progenitor masses should be different by more than 5\%.
If the two masses are close enough, the binary will explode nearly
at the same time leaving the neutron star$-$neutron star (ns,ns) binary.
In this case, there is
not enough time to change neutron star into black hole by accretion.}
first explodes leaving a neutron in the core. If the binary is within the
radius of the companion in the giant stage, the first born neutron star
will spiral into the envelope of the expanding companion in the giant stage.
We assume the envelope of the giant to be convective.

In the rest frame of the compact object, Bondi-Hoyle-Lyttleton accretion
of the hydrogen envelope of density $\rho_\infty$ and velocity $V$ is
   \be
   \dot M=2.23\times 10^{29} \ \frac{M_{co,1}^2}{V_8^{3}} \ \rho_\infty\
   {\rm g\ s^{-1}}
   \ee
where $M_{co,1}\equiv M_{co}/1\msun$
is the mass of the compact object, $V_8$ is the
velocity in unit of 1000 km s$^{-1}$, and $\rho_\infty$ is given in
g cm$^{-3}$. The density $\rho_\infty$ required for the hypercritical
accretion, $\dot M=10^8 \dot M_{Edd} \approx 10^{26}$ g s$^{-1}$, is
   \be
   \rho_\infty = 0.44\times 10^{-3} \frac{V_8^3}{M_{co,1}^{2}} \
   {\rm g \ cm^{-3}}.
   \ee
Using the Kepler law for circular orbits $V^2=GM_{tot}/a$, where $M_{tot}$ is
the combined mass of the compact object and the stellar material interior
to the orbit of the compact object and $a$ is the binary separation,
one finds
   \be
   \rho_\infty = 2.1\times 10^{-5} \frac{1}{M_{co,1}^{2}}
   \left(\frac{M_{tot,10}}{a_{12}}\right)^{3/2} {\rm g\ cm^{-3}},
   \ee
where $M_{tot,10}=M_{tot}/10^{10}M_{\odot}$ and $a_{12}=a/10^{12}cm$.
Since this critical density is of the same order as
the average density of the hydrogen envelope of the giant star,
the hypercritical accretion $\dot M\gsim 10^8 \dot M_{Edd}$
can occur in this spiral-in process. This rapid accretion can add $\sim 1\msun$
to the first-born neutron star in less than about one year, which is
enough to turn the
first born neutron star into a black hole. In this process, the orbital
radius shrinks by a factor $\sim 50$.

If the companion is massive enough to have a supernova explosion,
the binary system
will end up as a black hole$-$neutron star (bh,ns) binary instead of a
(ns,ns) binary. If the companion is less massive,
the binary will end up as black hole$-$CO white dwarf (bh,cowd) binaries.
Brown, Lee, Portegies Zwart, \& Bethe (1999)
show that there are big discrepancies between
the predicted and the observed (ns,ns) and (ns,cowd)$_{c}$ (i.e. with
circular orbits) binary systems. They solved this discrepancy by arguing that
in both systems the first
born neutron star turned into black holes by accretion
leaving behind (bh,ns) and (bh,cowd) binaries. The (bh,ns) binaries are
one of the most important sources for gravitational detectors.
The chirp mass of double neutron star binary with masses $1.4\msun$
is $M_{chirp}=\mu^{0.6} (M_1+M_2)^{0.4}=1.2\msun$, while that of
the (bh,ns) binary with $M_{BH}=2.4\msun$ is $M_{chirp}=1.6\msun$.
The (bh,ns) binaries, if they exist, can be more easily detected
by factor $(1.6/1.2)^3=2.4$ than the (ns,ns) binaries.
Since the LIGO is based on one observation of (ns,ns) binary, their
unseen (bh,ns) will enhance the detectability of LIGO by factor of
$\sim$30 considering the effect of the chirp masses.
This will be tested in near future.

One of the most strong criticisms on the hypercritical accretion was
the angular momentum and possibility of a strong jet along the rotation axis.
Even thouth these remain as unsolved problems,
Brown, Lee, \& Bethe (2000) showed that
Narayan \& Yi (1994) ADAF solution can be applied in this case if the
accretion flows is radiation-dominated. With the hypercritical
accretion, the inner boundary of accretion flow can be extended to the
marginally bound orbit ($r_{mb}=2 R_{Sch}$ for Schwarzschild black holes).
Since $r_{mb}\approx 9$ km, the inner boundary of accretion disc can
be extended to the neutron star surface $r_{NS}\gsim 10$ km.
By using a reasonable viscosity parameter $\alpha=0.05$,
they found that the neutrino cooling in the disc
is almost negligible $<10^{-7}$ of viscosity generated energy.
Brown et al. (2000) argued that
in order to avoid strong polar jets or pile-up up of material above neutron
star surface, the neutrinos are generated at the neutron surface
to carry out the pressure and energy from the system.

%--------------------------------------------------------------------
\section{Hypercritical Accretion onto a Magnetar}
Despite a very strong magnetic field, the dynamic effects of the magnetic
field would be negligible if the mass accretion rate satisfies
   \be
   \dot M &>& \frac 12 B^2\left(\frac{GM}{R^5}\right)^{1/2}
    \sim  7\times 10^{25}\ B_{14}^2
        \left(\frac{M_{1.5}}{R_6^5}\right)^{1/2} {\rm g\ s^{-1}}
 \nonumber\\
   &\sim & 4\times 10^7 \ B_{14}^2 \left(\frac{M_{1.5}}{R_6^7}\right)^{1/2}
   {\dot M}_{\rm Edd}
   \ee
Such a possibility is realized when the accretion is hypercritical.

During hypercritical accretion, even when the electromagnetic dipole
radiation occurs for a fast spinning pulsar, the spin-up by hypercritical
accretion can counter the spin-down. The spin-down torque due to the
dipole radiation emission is
   \be
   N_{em}={2\over 3c^3} B^2 R^6 \sin^2\theta\ \Omega^3
   \ee
and the spin-up torque due to accretion is
   \be
   N_{acc}={\dot M} \tilde a\ (GMR)^{1/2}
   \ee
where $\tilde a\sim 0.5$ is the sub-Keplerian rotation factor
in the hypercritical accretion (Brown, Lee, \& Bethe 2000).
The equilibrium spin is reached at the neutron star spin frequency
   \be
   \Omega_\NS \approx 1.4\times 10^3
   \left(\frac{\tilde a_{0.5}\ {\dot M}_{26}\ M_{1.5}^{1/2}}
   {\sin^2\theta \ R_{6}^{11/2}\ B_{14}^{2}}\right)^{1/3}~s^{-1}.
   \ee
By assuming $\langle \sin^2\theta\rangle \sim 0.5$, one gets
$\Omega_{NS}\sim 1.8\times 10^3 ~{\rm s}^{-1}$ with other quantities
in unities.
The collapse of the neutron star to a black hole results in a rapidly
spinning black hole with a spin frequency
   \be
   \Omega_\H=\left(\frac{R_\NS}{R_\H}\right)^2
   \Omega_\NS\approx 4\Omega_\NS\sim 7.2\times 10^3\ s^{-1}
   \ee
which corresponds to black hole spin of $\tilde a\sim 0.1$.
The magnetic field strength
   \be
   B_\H= \left(\frac{R_\NS}{R_\H}\right)^2 B_\NS
   \approx 4B_\NS \approx 4\times 10^{14} B_{\NS,14}\ \G
   \ee
where we have assumed the angular momentum conservation and flux-freezing.

If the magnetar enters the hypercritical accretion regime as a slowly spinning
neutron star, the spin-up during the hypercritical accretion will take place
on a time scale
    \be
    t_{spin-up}\sim \frac{I_\NS \Omega_\NS}{N_{acc}}
    \sim 1.6\times 10^7~{\rm s} \sim 0.5~{\rm yr} %\frac{1}{a_{0.5}}
     %\left[\frac{\Omega_\NS}{\Omega_K(R_\NS)}\right]~s
    \ee
which is comparable to the spiral-in time scale ($\sim 1$ yr)
during the common envelope phase.

%---------------------------------------------------------------------
\section{Energetics of Blandford-Znajek Process}
Based on the steady pulsar periods and spin-down rates,
the magnetic field of pulsars are generally estimated by assuming the dipole
radiation induced by magnetic fields. The typical fresh neutron stars in
the pulsar island have $\sim 10^{12}$ G, and the recycled ones have
$\lsim 10^{10}$ G. The discoveries of the magnetar and AXP
suggest the possibilities of the strong magnetic field $\gsim 10^{14} G$.
Even though AXP and magnetars raised questions on the
origin of the pulsar mechanism,
we assume the dipole radiation as origin of the pulsar in these systems.
The probability of pulsar with strong magnetic field is assumed to be
$\sim 10\%$ (Kouveliotou et al. 1998, 1999).

After the discovery of GRB afterglows, the theoretical understandings
of the formation of GRBs have been improved quite a lot. However, the
mechanism of the central engine is still controversial.
The electromagnetic extraction of the BH rotating energy by the
Blandford-Znajek process (Bandford \& Znajek 1977) is one of
the candidate mechanisms, in which
the strong magnetic field is essential to power the GRB.
Typically $B_H \sim 10^{15}$ G can generate
enough energy to power the GRB from rotating black holes.
However, the existence of the strong magnetic field in the black holes
is uncertain because the strong magnetic fields diffuse in very short
time scales.

We suggest that the Blandford-Znajek process
can operate if the collapsing neutron stars,
with $B\gsim 10^{14}$ G, carry the magnetic fields with them,
amplifying the fields upto $\sim 10^{15}$ G.
In this scenario, the existence of neutron stars prior to the black holes
is essential.

The hypercritical accretion prior to the collapse to a
black hole spins up the neutron star and the black hole is expected
to be spinning at rotational angular velocity close to
$\tilde a\sim 0.1$.
Once black hole is formed, the spin-down torque due to the dipole
emission is not working, and the spin-up torque due to accretion
can increase the black hole spin further. So, our estimate given above
may be a low limit.

The Blandford-Znajek power (Lee, Wijers, \& Brown 2000a)
from  the rotating black hole with the
angular momentum
$J = a M$ and the magnetic field $B_H$ is given by
\be
P_{\rm BZ} &\sim& 10^{50}a^2 B_{H,15}^2 (M/M_{\odot})^2 {\rm ~erg\ s^{-1}}.
\ee
Using the values in the previous section, we have
$P_{\rm BZ} \sim 3.6\times 10^{47}$ erg s$^{-1}$.
The total available rotational energy is 0.06\% of the rest mass
$\sim 2\times 10^{51}$ erg (Lee, Wijers, \& Brown 2000a).

%---------------------------------------------------------------------
\section{The Role of Black Hole Formation in the Spiral-In Phase}

When the first born star with strong magnetic field accretes material
from the companion envelope, the
black hole should be formed inside the envelope. Since the black hole
is formed in the process of the hypercritical accretion, the temperature
near the surface of the neutron star should remain $\sim 1$ MeV
until the collapse into a black hole. This temperature is required to
generate neutrinos. In this case, the spin frequency for the
r-mode instability is so high  that we can neglect the effect
(Lindblom \& Owen 1999).
If the Blandford-Znajek process works at the time of black hole formation,
there is enough energy to generate GRBs. However, because of the optically
thick envelope, gamma rays cannot come out through the hydrogen (or helium)
envelope. Instead of GRBs, the energy will pile up in the envelope
to generate supernova. The main difference from the normal
SN scenarios is that the SN is not induced by the core collapse of the
progenitors. The natural consequence is the strong asymmetry in the
SN events. Since the companion mass is not the unique parameter, even the
less massive stars (normal white-dwarf progenitors) can have SN events
with high kick velocities.

The available energies powered by rotating black holes ($> 1.5 \msun$)
with Blandford-Znajek process can be
$\sim 10^{51}$ erg. Since the binding energies ($0.6 GM^2/R$)
of the hydrogen envelope of giant stars ($10\msun$)
are $\sim {\cal O}(10^{48})$ erg,
newly formed black holes have enough energy to blow the envelope off.
If only $1\%$ of this SN energy is available for the kinetic energy of
the He core of the giant star, the available kick velocities are
$\sim {\cal O} (1000$ km s$^{-1}$).

Fast moving white dwarfs from SN remnants, if exists,
will be the consequences of this "companion induced SN events".
If the black holes are formed just before the normal SN events of the
companions, the double SN are also possible leaving (bh,ns) as remnants.
Especially, in the later case, the neutron star can have double kicks,
resulting in very high kick velocities, $>1000$ km s$^{-1}$.

%---------------------------
Bethe \& Brown (1998) argued that the formation rate of $(bh,ns)$
binaries is $10^{-4}$ yr$^{-1}$ per galaxy, where the low mass black holes
($\sim 2.5\msun$) are formed in the common envelope evolution.
By assuming the flat $q$
distribution, Brown, Lee, Portegies Zwart, \& Bethe (1999) also argued that
the formation rate of circular $(bh, cowd)_{c}$ binaries
is $\sim 1.7\times 10^{-4}$ yr$^{-1}$ per galaxy.
In their scenario the first born neutron star went into the common
envelope evolution, where the neutron star is converted into black hole by
hypercritical accretion. According to our scenario, in the formation
of $(bh,ns)$ and $(bh,cowd)_{c}$ binaries by hypercritical accretion,
about $10\%$ of the
neutron stars have very strong magnetic field and will have the companion
(newly formed black holes) induced supernova-like explosions.
As a result $\sim 10\%$ of progenitors of $(bh,cowd)_{c}$ binaries
end up as eccentric $(bh,cowd)_{e}$ binaries instead of circular
ones.
Also $\sim 10\%$ of progenitors of $(bh,ns)$ binaries
have double explosions with two different centers of explosion.
   % \footnote{
   % Some of $(bh,ns)$ progenitors with strong magnetic fields
   % also may evolve into eccentric $(bh,cowd)_{e}$ binaries.
   % }
By assuming that half of the progenitors of $(bh,cowd)_{e}$ or $(bh,ns)$
binaries survive the explosion induced by rotating black holes,
we have  the formation rates of
eccentric $(bh,cowd)_e$ binaries or double explosions
   \be
   R\sim 10^{-5}\ {\rm yr}^{-1} {\rm \ per \ galaxy}.
   \label{eq.R}
   \ee
Since the cooling time of CO white dwarfs,
i.e. the time required to reach luminosities
$\log (L/L_\odot)=-4.5$ with $L_\odot=$ solar luminosity, is
$1-10$ Gyr (Salaris et al. 1997), the number of eccentric $(bh,cowd)_e$
binaries with luminosity $\log(L/L_\odot)>-4.5$ is
   \be
   N \sim 10^4-10^5 {\rm \ in  \ Galaxy}.
   \ee
The number of non-pulsating eccentric $(ns,cowd)_e$ binaries
with luminosity $\log(L/L_\odot)>-4.5$ is $\sim 10$ times
more popular than $(bh,cowd)_e$ systems. In order to identify
$(bh,cowd)_e$ systems, one need
well-established maximum neutron star masses.
If there are mechanisms to identify the black holes from dead pulsars
in addition to the mass (Balberg et al. 1999), one can detect
$(bh,cowd)_e$ systems.

%---------------------------------------------------------------------
\section{Discussion}
We have looked into a possibility that a supernova-like explosion could occur
during the common envelope evolution of a strongly magnetized neutron
star. The "magnetar" accretes in the form of the hypercritical accretion
during the spiral-in phase and collapses to a rapidly spinning black hole
with a strong magnetic field near the horizon. Then the explosion results from
the sudden release of energy through the Blandford-Znajek mechanism. In this
scenario, the strong magnetic field naturally occurs as it is derived from the
strong field of the pre-collapse "magnetar".

There are a number of points to be clarified in the proposed scenario some of
which could be observationally testable.
First, if the magnetic field of the magnetar decays on a time scale
$\sim 10^3-10^4$
yrs as is often argued for the evolutionary behavior of magnetars and anomalous
X-ray pulsars (Colpi, Geppert, \& Page 2000),
then the common envelope phase has to follow the magnetar
formation. Second, during the accretion phase, most of the emitted energy
(including the dipole radiation) is ultimately radiated from the neutron star
in the form of the neutrinos. Third, if the Blandford-Znajek phase occurs
while the neutron star collapses into black hole in the envelope,
the resulting off-center explosion
will give a high eccentricity, which should be testable in terms of the highly
eccentric binary systems. Fourth, there will be little metal spread from the
core since there will be no core collapse. The dominant effect would be
blowing-off the hydrogen envelope preceded by the expansion of the envelope.

If the merger of the companion core with the neutron star occurs before
the black hole formation, then the explosion will not cause an off-center
explosion. Although this possibility exists, the estimated accretion time
scale on which the neutron star can accrete enough mass to collapse to
a black hole is comparable to the spiral-in time scale. This implies that the
off-center explosion is possible.

%--------------------------------------------------------
\acknowledgements

CHL is partly supported by the U.S. Department of Energy under grant
DE-FG02-88ER40388. IY is supported in part by 1999-2000 KIAS Research Fund and
KRF Research Fund KRF 1998-001-D00365. HKL is supported in part by BK21 Program
 of Ministry of Education and by KOSEF Grant No. 1999-2-112-003-5.
%--------------------------------------------------------
%
%
%--------------------------------------------------------

\end{document}